\documentclass[aps,pra,twocolumn,superscriptaddress,groupedaddress,showpacs]{revtex4-1}  
\usepackage{graphicx}  
\usepackage{dcolumn}   
\usepackage{bm}        
\usepackage{color}
\usepackage{float}
\usepackage{times}
\usepackage{amsfonts, amsmath, amsthm, amssymb} 
\usepackage{adjustbox}
\usepackage[colorlinks=true,citecolor=blue,linkcolor=blue,urlcolor=blue]{hyperref}
\usepackage{color}

\begin{document}
	\title{Enhanced Quantum Synchronization via Quantum Machine Learning}
	\author{F. A. C{\'a}rdenas-L{\'o}pez}
	\email[Corresponding authors:~~]{\\francisco.cardenas@usach.cl}
	\affiliation{Departamento de F\'isica, Universidad de Santiago de Chile (USACH), Avenida Ecuador 3493, 9170124, Santiago, Chile}
	\affiliation{Center for the Development of Nanoscience and Nanotechnology 9170124, Estaci\'on Central, Santiago, Chile}
	\author{M. Sanz}
	\email[]{mikel.sanz@ehu.es\\}
	\affiliation{Department of Physical Chemistry, University of the Basque Country UPV/EHU, Apartado 644, 48080 Bilbao, Spain}
	\author{J. C. Retamal}
	\affiliation{Departamento de F\'isica, Universidad de Santiago de Chile (USACH), Avenida Ecuador 3493, 9170124, Santiago, Chile}
	\affiliation{Center for the Development of Nanoscience and Nanotechnology 9170124, Estaci\'on Central, Santiago, Chile}
	
	\author{E. Solano}
	\affiliation{Department of Physical Chemistry, University of the Basque Country UPV/EHU, Apartado 644, 48080 Bilbao, Spain}
	\affiliation{IKERBASQUE, Basque Foundation for Science, Maria D\'iaz de Haro 3, 48013 Bilbao, Spain}
	\affiliation{Department of Physics, Shanghai University, 200444 Shanghai, People's Republic of China}

	\begin{abstract}
		We study the quantum synchronization between a pair of two-level systems inside two coupled cavities. By using a digital-analog decomposition of the master equation that rules the system dynamics, we show that this approach leads to quantum synchronization between both two-level systems. Moreover, we can identify in this digital-analog block decomposition the fundamental elements of a quantum machine learning protocol, in which the agent and the environment (learning units) interact through a mediating system, namely, the register. If we can additionally equip this algorithm with a classical feedback mechanism, which consists of projective measurements in the register, reinitialization of the register state and local conditional operations on the agent and environment subspace, a powerful and flexible quantum machine learning protocol emerges. Indeed, numerical simulations show that this protocol enhances the synchronization process, even when every subsystem experience different loss/decoherence mechanisms, and give us the flexibility to choose the synchronization state. Finally, we propose an implementation based on current technologies in superconducting circuits. 
	\end{abstract}
	
	\pacs{Machine Learning, Quantum Information processing, Quantum Synchronization.}
	\maketitle
	\section{Introduction}
	Synchronization phenomenon refers to a set of two or more self-sustained oscillators with different frequencies that are forced to oscillate with a common effective frequency as result of the weak interaction between the system component \cite{1}, which is characterized by its non-reversible behaviour due to the time-reversal symmetry breaking \cite{2}. This phenomenon has been observed and used in biological systems \cite{3,4}, engineering \cite{5}, and geolocalization, to name a few. During the last decade, significant progress has been made in the development of quantum platforms such as trapped ions \cite{6,7}, nanomechanical resonator \cite{8,9,10} as well as superconducting circuit and circuit quantum electrodynamics (cQED) \cite{11,12,13}. This important progress has made possible to study the synchronization phenomena at the quantum level \cite{14,15,16,17,18,19,20,21,22}. Initially, arrays of quantum harmonic oscillators were studied. These systems have a classical limit since they can be effectively treated as classical systems when the oscillators are highly, allowing a natural comparison between classical and quantum synchronization. However, the study of synchronization in quantum systems without a classical counterpart such as two-level systems becomes non-trivial and controversial. It has to be studied, among other techniques, through the natural observables of these systems \cite{21,22,23,24,25,26,27,28}.\par
	Artificial intelligence (AI) and machine learning (ML) have concentrated much attention in the last years. ML consists of adaptive computational algorithms which can improve their performance, employing the history of data record \cite{29}. In physics, several fields have profited from the advantages offered by ML, such as material science \cite{30}, high energy physics \cite{31} and condensed matter physics \cite{32}, finances \cite{33}, state discrimination \cite{34}, design efficiently experiments \cite{35}, just to name a few. Essentially, there are three types of learning in ML, namely, supervised learning, unsupervised learning and reinforcement learning \cite{36}. In supervised learning, the system learns from initial data to make future decisions. Regression (continuous output) and classification (discrete output) are considered as the archetypical supervised learning algorithm. In unsupervised learning, the classes are not defined from the beginning (classification), but they naturally emerge from the initial data. In other words, the data is organized in subsets based on correlations found by the algorithm. Data clustering is the most usual example of the unsupervised learning algorithm. In reinforcement learning \cite{37}, there is a scalar parameter, named rewarding, which evaluates the performance of the learning process. Depending on the rewarding, the system can decide whether the learning process is optimized or not.\par
	The use of quantum algorithms to accomplish machine learning tasks as well as the use of machine learning algorithms to solve quantum information processing tasks has led to the emergence of Quantum machine learning \cite{38,39,40,41,42,43}. In this field, arise a new paradigm concerning the nature of the machine learning components, namely, the agent and the environment \cite{44}. Four categories related to the nature of the learning components can be identified in this two-party system: classical-classical (CC), classical-quantum (CQ), quantum-classical (QC) and quantum-quantum (QQ). The first of them is related to the classical machine learning. The second corresponds to the case where classical machine learning can address quantum tasks. The third corresponds to the quantum variance of classical machine learning algorithms. In this case, the quantum algorithms which overtake the performance of their classical counterpart have already been shown in supervised and unsupervised machine learning \cite{44,45,46,47,48,49,50}. The last category corresponds to the case in which quantum systems comprise both agent and environment. In such a case, the definition of learning has not been explicitly defined and has to be interpreted as the optimization of certain figure-of-merit \cite{43,52,53}. Recently, a novel perspective of using ML algorithms to enhance quantum tasks has emerged, such as the use of genetic algorithms to reduce errors in quantum gates or quantum simulations \cite{48}, to learn an unknown transformation [54] or speed up quantum tomography \cite{55}, to generate a quantum adder \cite{56}, or to construct a quantum autoencoder \cite{57} Furthermore, the use of neural network has proven to be a useful tool to address many-body physics problems, to solve variationally the time-independent and time-dependent Schrödinger equation \cite{32}, to perform quantum state tomography \cite{55} and to classify phases of matter \cite{58}.\par
	In this article, we address the following question: can we understand synchronization phenomena as a machine learning protocol? To answer this question, we rely on the digitalization of the master equation governing the system dynamics. We show that the digitized dynamics leads to the same result obtained in the analog case. We can identify in this digitalization, fundamental elements of a quantum machine learning (QML) protocol \cite{54}. We realize the presence of QQ paradigm of QML. We find that synchronization of two qubits, when implemented as a QML protocol, can be enhanced by adding a feedback mechanism, which lead to a reinforcement learning protocol. The enhancement on the synchronization relies on the increase of the mutual information in the Agent-Environment subsystem. Furthermore, the application of the classical feedback protocol induces complete synchronization of two-level observables in dynamics which do not synchronize in absence of feedback mechanism. Finally, we propose an implementation with current technology in superconducting qubits.\par
	
	This article is organized as follows: In section \ref{sec2}, we present the digital-analog decomposition of the master equation governing the system dynamics and show that this decomposition yields quantum synchronization between a pair of two-level systems. In section \ref{sec3}, we show that it is possible to enhance the synchronization by adding a classical feedback mechanism in the proposed decomposition. The enhancement is quantified through the quantum mutual information finding conditions under which synchronization arises. Besides, we compute the expectation value of the two-level observables for the Agent and Environment subsystem, showing that the complete set of two-level observables oscillates with a common frequency. In section \ref{sec4}, we discuss the implementation of the ML protocol in superconducting circuits, considering the near-term available technologies. In section \ref{sec5}, we discuss the numerical aspects of the digital-analog decomposition of the master equation governing the system dynamics. Finally, in section \ref{sec6}, we present conclusions, and perspectives.
	
	\section{Digitized quantum synchronization}\label{sec2}
	\begin{figure}[b]
		\centering
		\includegraphics[width=1\linewidth]{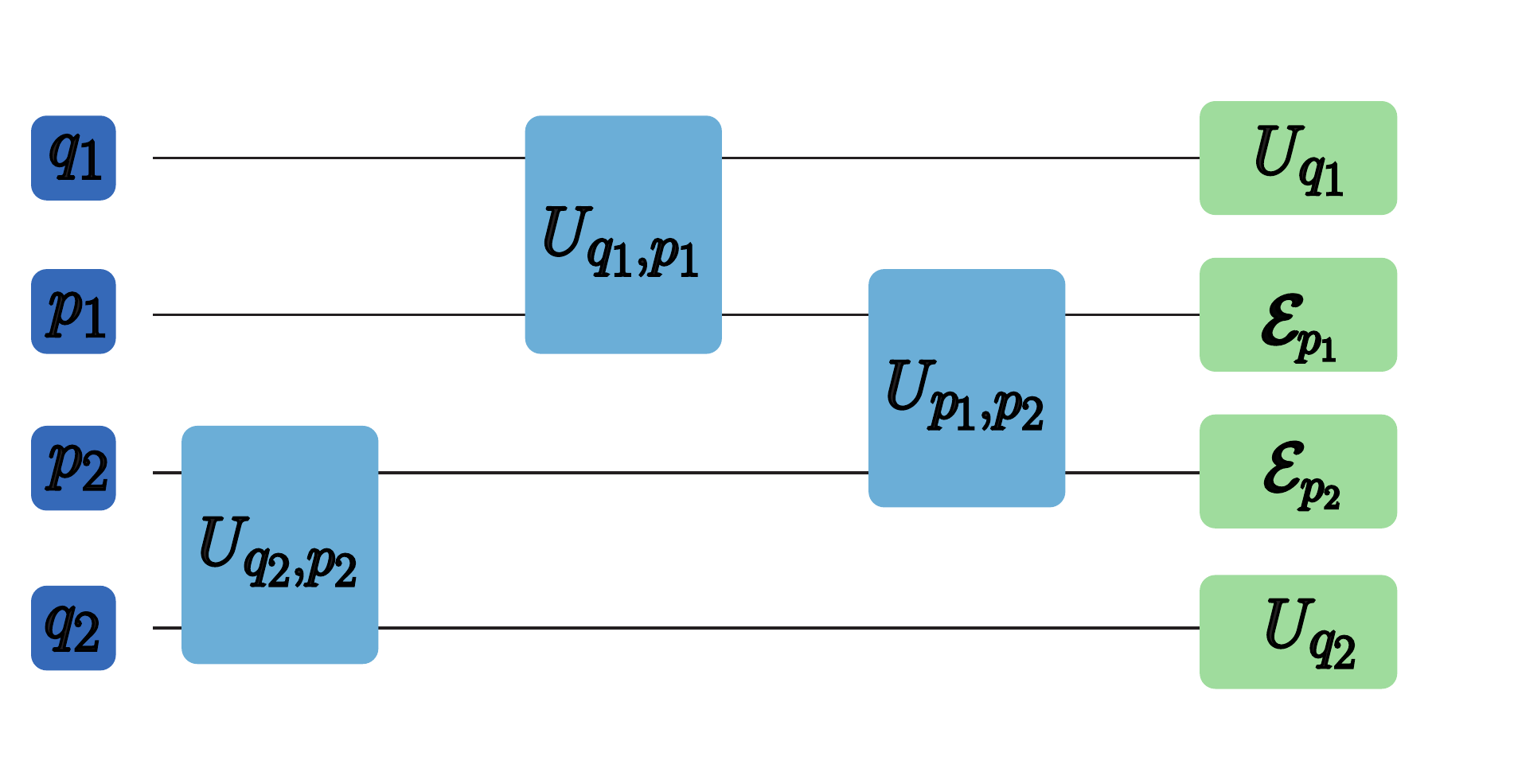}
		\caption{Schematic diagram for the digitized master equation, where $q_\ell$ and $p_\ell$ ($\ell=1,2$) represent the qubits and the cavities composing the system, respectively. The purple box corresponds to the global gates represented by $U_{q_\ell,p_\ell}$ and $U_{p_1,p_2}$. The first of them stands for the qubit-cavity interaction (Jaynes-cummings terms), whereas, the second gate is the cavity-cavity interaction (hopping term). Besides, the green boxes correspond to the local gates, which can be divided into two types: the local gate $U_{q_\ell}$ is the free evolution of the qubits, and $\mathcal{E}_{p_{\ell}}$ is the non-unitary dynamics for the cavities represented by the dynamical map $\mathcal{E}$.}
		\label{fig:fig1}
	\end{figure}
	\begin{figure}[!t]
		\centering
		\includegraphics[width=1\linewidth]{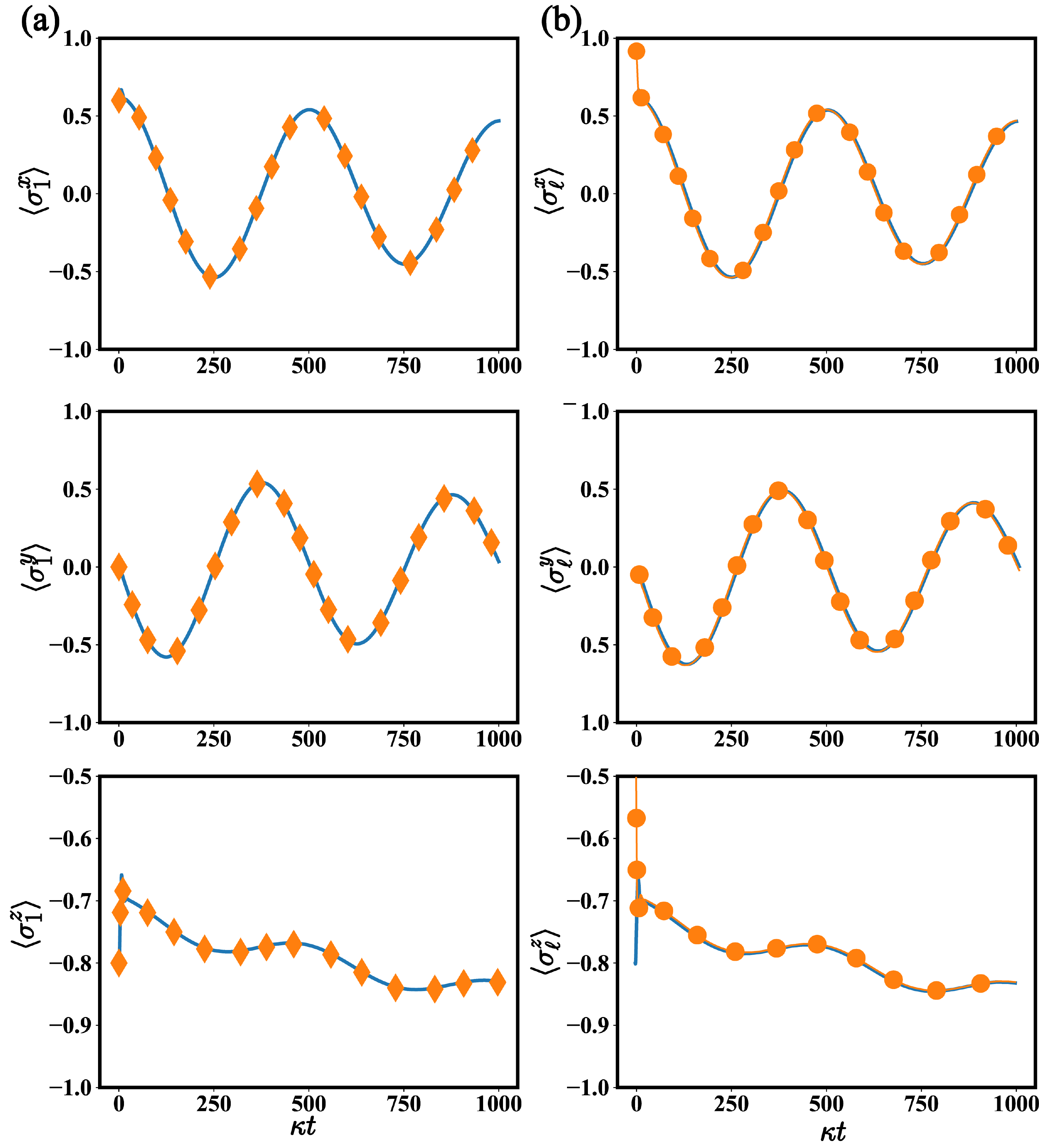}
		\caption{Time evolution for the average for the qubit's observables $q_1$. Continuous blue line stands for the mean values obtained by solving directly the master equation on Equation (\ref{Eq1}). The orange dotted line corresponds to the same mean values obtained through the digital-analog decomposition of the master equation shown in Figure (\ref{fig:fig1}). The system parameters in this case are $\Delta_1=\Delta_2=J=10$, $\delta_1=\delta_2=0$, $g=2\kappa$ and $\Omega=5\times{10}^{-4}$. The initial state of the system is $|\Psi(0)\rangle=(\sqrt{0.9} |g\rangle+ \sqrt{0.1}|e\rangle)\otimes(\sqrt{0.7}|g\rangle+ \sqrt{0.3}|e\rangle)\otimes|0\rangle|0\rangle$, where $|e\rangle(|g\rangle)$ stands for the excited (ground) state of the qubits and $|0\rangle$ is the vacuum state of the cavity}
		\label{fig:fig2}
	\end{figure}
	Let us consider a system composed of two dissipative cavities containing each one a two-level system. Both cavities interact via a hoping interaction, and a coherent driving field acts in one of the two-level systems to counterbalance the dissipation present in both cavities. The dynamics of the system is described by the master equation ($\hbar=1$). 
	\begin{eqnarray}
	\label{Eq1}
	\dot{\rho}(t) = -i[\mathcal{H},\rho] + \kappa\sum_{\ell=1}^{N=2}(2a_{\ell}\rho a_{\ell}^{\dag} - a_{\ell}^{\dag}a_{\ell}\rho - \rho a_{\ell}^{\dag}a_{\ell}),
	\end{eqnarray}
	where the Hamiltonian $\mathcal{H}$ is expressed in the rotating frame with respect to the laser field as
	\begin{eqnarray}\nonumber
	\label{Eq2}
	\mathcal{H}&=&\sum_{\ell=1}^{N=2}\bigg[\Delta_{\ell}a^{\dag}_{\ell}a_{\ell}+ \frac{\delta_{\ell}}{2}\sigma^{z}_{\ell} + g_{\ell}(a_{\ell}^{\dag}\sigma^{-}_{\ell} + a_{\ell}\sigma^{+}_{\ell})\bigg]\\ &+&J\left(a_{1}^{\dag}a_2+a_1a_{2}^{\dag}\right)+\Omega\sigma_1^x.
	\end{eqnarray} 
	Here, $a_\ell^\dag\left(a_\ell\right)$ is the creation (annihilation) boson operator of the lth field mode, while $\sigma_\ell^k$ stands for the $k$-component Pauli matrix. $\Delta_\ell=\omega_{p,\ell}-\omega_d$ is the detuning between the lth field mode $\omega_{p,\ell}$ and the driving frequency $\omega_d$, while $\delta_\ell=\omega_{q,\ell}-\omega_d$ stands for the detuning between the lth qubit frequency $\omega_{q,\ell}$ and the driving frequency. Moreover, $g$ is the coupling strength between the field mode and the two-level system. Finally, $\Omega$ corresponds to the strength of the driving field, J is the coupling strength between cavities, and $\kappa$ is the decay rate for the cavities.\par
	
	The master equation in Equation (\ref{Eq1}) works in the bad-cavity regime. In this regime, it is possible to define a hierarchy of time scales as follows: $\{\kappa\ll g_\ell \ll \delta_{\ell}\}$. Decoherence and energy dissipation of the two-level system is not considered because times involving these processes are too slow compared with, for example, the cavity dissipation (${\kappa}$) and the coherent coupling between the field modes and the two-level systems ($g_\ell$) \cite{59}. This type of regime can be achievable in quantum platforms such as trapped ions, cavity quantum electrodynamics, where the decay rates on the two-level systems are in the second range \cite{60} and superconducting circuits, where stables artificial atoms based on 3D architecture has proven to achieved coherence times at the millisecond scale \cite{61}. Therefore, the fundamental timescale to study the synchronization phenomena corresponds to photons leaking rate $\kappa$. In this system configuration, it is already proven that quantum synchronization between observables of two qubits is achieved \cite{62}.
	
	Our approach to this problem is to analyse the quantum synchronization between the two qubits by considering a digital-analog version of the master equation given in Equation (\ref{Eq1}). A digital-analog protocol is a hybrid approach to quantum computing that makes use of an analog entangling Hamiltonian, which provides the robustness, together with fast digital rotations, which provide the flexibility. This approach was proven to be universal \cite{63} and has been recently proposed as a possible alternative to the purely digital approach with quantum error correction in the NIST era. We show that both digital-analog simulation and fully analog simulation yield quantum synchronization. The decomposition of the master equation into digital-analog steps is shown in Figure (\ref{fig:fig1}). We can discriminate between the analog blocks (blue boxes) and digital blocks (green boxes). Analog blocks are associated with the interaction terms in Hamiltonian in Equation (\ref{Eq2}), which correspond to Jaynes-Cummings and hopping terms. The unitary operations can implement the dynamics associated with these terms $U_{q_\ell,p_\ell}$ ($\ell=\ 1,2$) and $U_{p_1,p_2}$ defined as
	
	\begin{subequations}
		\label{Eq3}
		\begin{eqnarray}
		{U}_{q_{\ell},p_{\ell}}&=&e^{-ig(a_{\ell}^{\dag}\sigma^{-}_{\ell} + a_{\ell}\sigma^{+}_{\ell})\Delta t}.\\
		{U}_{p_{1},p_{2}}&=&{e}^{-{iJ}\left(a_{1}^{\dag}a_{2}+a_{1}a_{2}^{\dag}\right)\Delta t}.
		\end{eqnarray}
	\end{subequations}
	On the other hand, we associate digital gates with the unitary dynamics of the free Hamiltonian in Equation (\ref{Eq3}), and the dissipative dynamics of the Lindbladian terms of the master equation in Equation (\ref{Eq1}). For both qubits, local gates only correspond to the evolution of their respective free Hamiltonians,
	\begin{subequations}
		\label{Eq4}
		\begin{eqnarray}
		{U}_{q_1}&=&e^{-i(\delta_{1}\sigma^{z}_{1}/2 + \Omega \sigma_{1}^{x})\Delta t},\\
		{U}_{q_2}&=&e^{-i\delta_{2}\sigma_{2}^{z}\Delta t/2},
		\end{eqnarray}
	\end{subequations}
	for the cavity, ${p}_\ell$, the local operation is represented by the dynamical map $\mathcal{E}_{p_\ell}$, which can be written as the following master equation \cite{64}
	\begin{eqnarray}
	\label{Eq5}
	\dot{{\rho}}\left({t}\right)=\sum_{\ell=1}^{N=2}{-{i}\left[\Delta_{\ell}{a}_{\ell}^\dag{a}_{\ell},{\rho}\right]+{\kappa}\left(2{a}_{\ell}{\rho}{a}_{\ell}^\dag-\{{a}_{\ell}^\dag{a}_{\ell},{\rho}\}\right)}. 
	\end{eqnarray}
	
	for a time $\Delta t$. As we mentioned previously, the cavities are operating on the bad-cavity regime. Therefore, the fundamental time scale of the system is the leak of photons $\kappa$. Thus, the gate time will be subdivisions of this scale. In section 5, we will discuss the numerical analysis about the convergence and the gate decomposition of the proposed digitalization.
	
	We compute the expectation value of the observables of a two-level system, namely, $\{{\sigma}^{x},{\sigma}^{y},{\sigma}^{z}\}$ of ${q}_\mathbf{1}$ by using both method, i.e., by solving directly the master equation given in Equation (\ref{Eq1}) and the proposed digital-analog decomposition and compare them. In Figure (\ref{fig:fig2}a), it is shown that, for a given time decomposition ($\kappa t$ divided into 100 slides), the proposed decomposition leads to the same result obtained by directly solving the master equation. Likewise, to demonstrate that the digital-analog approach leads to synchronization between the pair of two-level systems, we also compute the expectation values of both two-level systems (${q}_\mathbf{1}$ and ${q}_\mathbf{2}$) through the aforementioned approach. Figure (\ref{fig:fig2}b) shows that, in this approach, the synchronization phenomenon is also achieved.
	
	We can interpret the proposed digitalized terms of the master equation as a machine learning protocol in which qubits learn from each other. Indeed, in a recent proposal for reinforcement quantum learning \cite{53,57}, the agent and the environment do not directly interact, and the learning process is mediated by an ancillary system, namely the register. In our setup, the synchronization between both qubits is similarly carried out through the field modes. In this case, we can identify the agent and the environment with the qubit ${q}_{1}$ and ${q}_{2}$, respectively, while the register is identified with the field modes. The novelty is that the register is now connected to a decoherence channel. In our case, the interactions among the elements are given by the digital-analog blocks. Thus, the digitalized master equation can be considered as a machine learning protocol without feedback.
	
	\section{Enhanced quantum synchronization}\label{sec3}
	\begin{figure}[t!]
		\centering
		\includegraphics[width=1\linewidth]{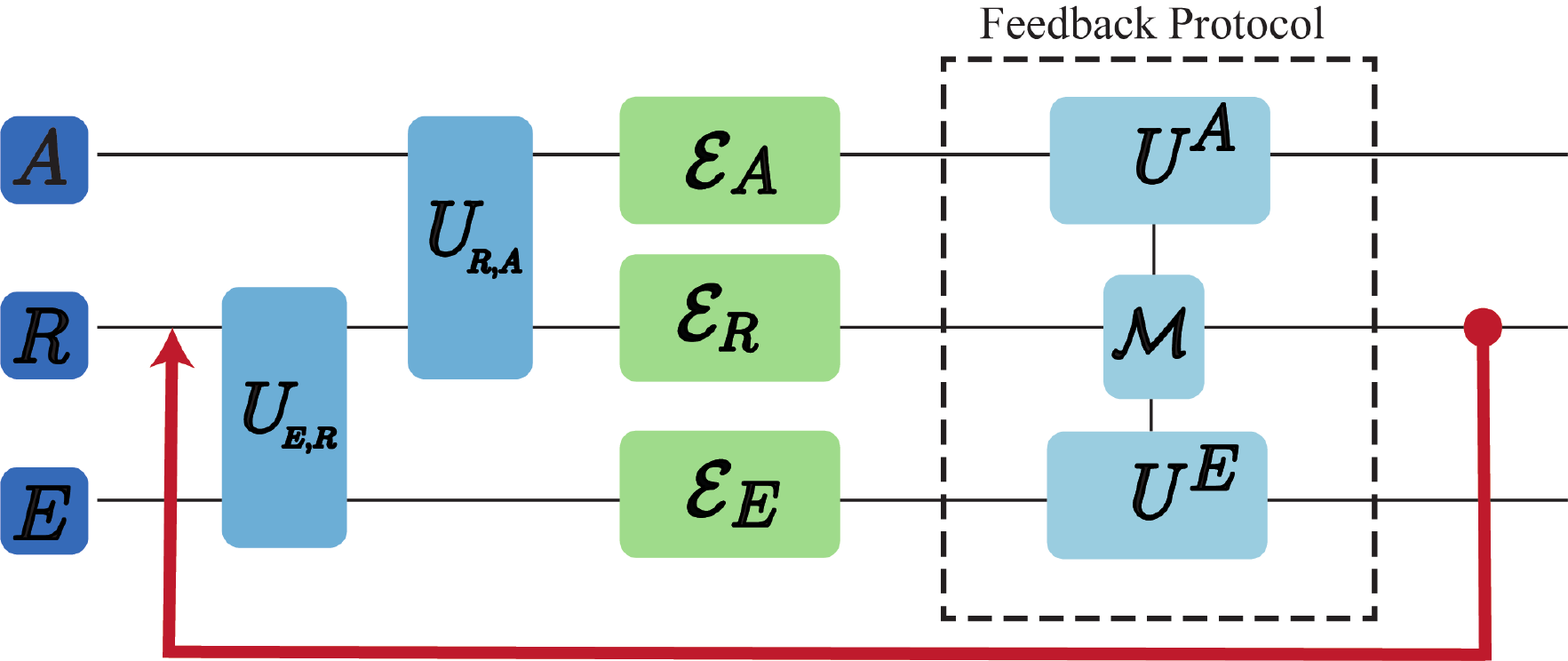}
		\caption{Schematic representation of the quantum circuit for a machine learning protocol with feedback. Here $A$, $E$ and $R$ correspond to the agent, environment and agent subsystems composed by qubits, respectively. Gates $U_{E,R}$ and $U_{R,A}$, act on the subspace formed by the agent (environment) and the register, respectively. $\mathcal{E}_\ell$ ($\ell=A, R, E$) are the non-unitary dynamics of the agent, qubit register and environment. The dashed box on the circuit represents the feedback process which is composed by a projective measurement $\mathcal{M}$ on the register. Depending on the measurement outcome the system is rewarded/punished through local operations $U_\ell$ ($\ell=A$, $E$) representing local gates acting on both agent and environment subspaces. The red arrow corresponds to the reinitialization of the register state.}
		\label{fig:fig3}
	\end{figure}
	In this section, we consider the synchronization process of two qubits as a quantum machine learning protocol equipped with feedback. We find that, by including a classical feedback protocol, it is possible to enhance the synchronization. Enhancement, in this case, means that, by analyzing some figure of merit, the protocol with feedback improves its performance over certain conditions. 
	
	Let us consider the scenario where the learning units, namely, agent (A), environment (E) and the register (R) are composed of two-level systems. Here, both agent and environment interact indirectly between each other through the ancillary system (register). This situation mimics the previous system because, in the bad-cavity limit, both cavities contain at most one excitation. As a result, the coupled cavities system can be well described by a two-level system.
	
	We characterized the quantum machine learning protocol by the set $\{S,A,r,V\}$ \cite{65}, where $S$ is the state space, represented by the Hilbert space expanding the complete system. $A$ is the Action-Percept set corresponding to the gates acting on the learning units. Furthermore, $r$ is the reward function that in our case correspond to a probability associated with a measurement outcome in the ancillary system. Finally, $V$ is our objective function, which must quantify the success and failure of our protocol. For instance, as we are interested in synchronization between a pair of two-level systems, an adequate figure of merit is the quantum mutual information \cite{27,28}
	
	\begin{eqnarray}
	\mathcal{I}={S}\left({\rho}_{A}\right)+{S}\left({\rho}_{E}\right)-{S}\left({\rho}_{{AE}}\right),
	\end{eqnarray}
	where $S(\rho)=-\rm{Tr}[\rho\log\rho]$ is the quantum von neuman entropy. In addition, $\rho_{A}$, $\rho_{E}$ and $\rho_{AE}$ are the reduced density matrix for the agen (A), environment (E) and agent-environment (AE) subsystem, respectively. The ue of quantum mutual information to quantify whether the system is sincronized or not relies on the fact that in two synchronized systems, the mutual information of this system is larger than zero. This fact was clearly explained in the Ref \cite{27,28}, where a set of two or more slightly different clocks are forced to oscillates with the same common frequency due to the collective interaction. In this case, the position of each clock is lost (high local entropy). Nevertheless, by knowing the state of one clock, it is possible to also know the state of the entire system (low global entropy). This scenario leads to the mutual information be different from zero. The other reason to consider this quantity as a figure-of-merit is that there is a simple relationship between mutual information in classical and quantum realms, respectively. For instances, in classical systems such as two coupled Van der Pool oscillators, $\mathcal{I}$ can be computed with the Shannon entropy, whereas, in quantum systems, the mutual information can be obtained through the von Neumann entropy \cite{27,28}.
	
	The quantum machine learning protocol for this situation is depicted in Figure (\ref{fig:fig3}). The first stage in our protocol is to perform the Action, i.e. transferring information from the environment qubit and encode it in the register. This action is done by the following analog block represented by the gate (light blue box in Figure (\ref{fig:fig3}))
	\begin{eqnarray}
	{U}_{{E},{R}}={e}^{{i}{J}_\mathbf{1}\left({\sigma}_{R}^+{\sigma}_{E}^-+{\sigma}_{R}^-{\sigma}_{E}^+\right) \Delta t},
	\end{eqnarray}
	where $\sigma^\pm$ is the ladder operator describing each two-level system, $J_{1}$ is the coupling strength between both subsystems and $\Delta t$ is the time in which the gate is acting. Afterwards, we perform the Percept corresponding to transfer the encoded information in the register subsystem towards the agent. Like the Action case, we represent the Percept by an analog block represented by the gate (light blue box in Figure (\ref{fig:fig3}))
	\begin{eqnarray}
	{U}_{{R},{A}}={e}^{{i}{J}_\mathbf{2}\left({\sigma}_{R}^+{\sigma}_{A}^-+{\sigma}_{R}^-{\sigma}_{A}^+\right) \Delta t} 
	\end{eqnarray}
	where $J_{2}$ is the coupling strength between both subsystems and $\Delta t$ is the time in which the gate is acting. The local operations acting on each subsystem are represented by the dynamical map $\mathcal{E}_{l}$, which can be represented by the following master equation for a time $\Delta t$ (green boxes in Figure (\ref{fig:fig3})),
	\begin{eqnarray}\nonumber
	\dot{{\rho}}\left({t}\right)&=&-{i}\left[\mathcal{H}_{q},{\rho}\right]+\sum_{\ell}{\gamma}\left(2{\sigma}_{\ell}^-{\rho}{\sigma}_{\ell}^+-\{{\sigma}_{\ell}^+{\sigma}_{\ell}^-,{\rho}\}\right)\ \\
	&+&\ \sum_{{\ell}}{{\gamma}_{\phi}\left(2{\sigma}_{\ell}^{z}{\rho}{\sigma}_{\ell}^{z}-{\rho}\right)}, 
	\end{eqnarray}
	where $\gamma$ corresponds to the two-level system relaxation rate, $\sigma_\ell^-$ to the ladder operator of the $\ell$th qubit, and $\gamma_\phi$ to the two-level system depolarizing noise rate. Finally, $\ell=\{A,\ E,\ R\}$ stands for the learning unit labels. Unlike the case studied in the last section, in this scenario, it is not possible to define an analogue to the bad-cavity limit even though the cavity in the previous section can be well described by a qubit. The reason of this assumption relies on the fact that the feedback mechanism requires systems with long coherences times. For example, in circuit quantum electrodynamics, digital coherent control on a superconducting circuits is within the microsecond scale, whereas, for near-terms devices, its coherence times is within the millisecond scale \cite{66,67}. Hence, the hierarchy in the time scale cannot be applied, and all subsystems must evolve under loss mechanisms. Finally, $\mathcal{H}_{q}$ is the Hamiltonian defined as
	\begin{eqnarray}
	\mathcal{H}_{q}=\frac{{\delta}_{A}}{2}{\sigma}_{A}^{z}+\frac{{\delta}_{E}}{2}{\sigma}_{E}^{z}+\frac{{\delta}_{R}}{2}{\sigma}_{R}^{z}+\Omega{\sigma}_{R}^{x}.
	\end{eqnarray}
	This Hamiltonian corresponds to the free energy of the agent, the environment and the driven register subsystem with a classical field. Notice that $\mathcal{H}_{q}$ has been expressed in the rotating frame with respect to the driving field. Here, $\sigma_{\ell}^{k}$ is the $k$-component Pauli matrix for the $l$th two-level system, $\sigma_{\ell}^\pm$ stands for the ladder operator for the $l$th qubit, $\delta_{\ell}=\omega_{q,\ell}-\omega_{d}$ is the detuning between the $\ell$th qubit and the driving field $\omega_{d}$, and $\Omega$ is the strength of the driving acting on the register.
	
	The next stage in our machine learning protocol is the application of the feedback mechanism. The feedback consists in measuring the register in the state $\left|{\psi}\right\rangle$. The probability associated to the measurement outcome constitutes our rewarding function $\mathcal{M}({\rho}_{{R}})_{|{\psi}\rangle}\in\{0,1\}$. Depending on the value taken by this function we decide whether we reward or punish the system. Rewarding here means that only the register is initialized in a new state, whereas punishment comprises local operations on the agent and environment learning units. For the rewarding, the register is measured in an eigenstate of ${\sigma}^{x}$ i.e., $\left|{\psi}\right\rangle=\left|\pm\right\rangle$. If the register is projected in the eigenstate $|+\rangle$, we initialize the register in the state $|-\rangle$, and we start with a new iteration. Otherwise, we initialize the register state in the state $|+\rangle$ and apply local operations on both agent and environment subsystem as punishment. These local operations correspond to
	\begin{eqnarray}
	{U}^{j}={e}^{-{i\pi}{\sigma}_{j}^{z}/2},
	\end{eqnarray}
	where $j=\{A,E\}$ is the index for the agent and environment. We find that the implementation of this machine learning protocol enhances the synchronization between a pair of two-level systems interacting with an ancillary system. The enhancement relies on the increase of the mutual information (correlations) between the agent and environment learning units. To accomplish that, let us consider the situation in which we initialize both agent and environment subsystems in orthogonal states, corresponding to the worse scenario for the learning process. Furthermore, we vary the qubit frequency of the agent $\delta_{A}$ and the coupling strength between the register-environment subsystems $J_{2}$. For this condition, we compute the mutual information after the system evolves up to a time $\kappa t=3$.
	\begin{figure}[t!]
		\centering
		\includegraphics[width=1\linewidth]{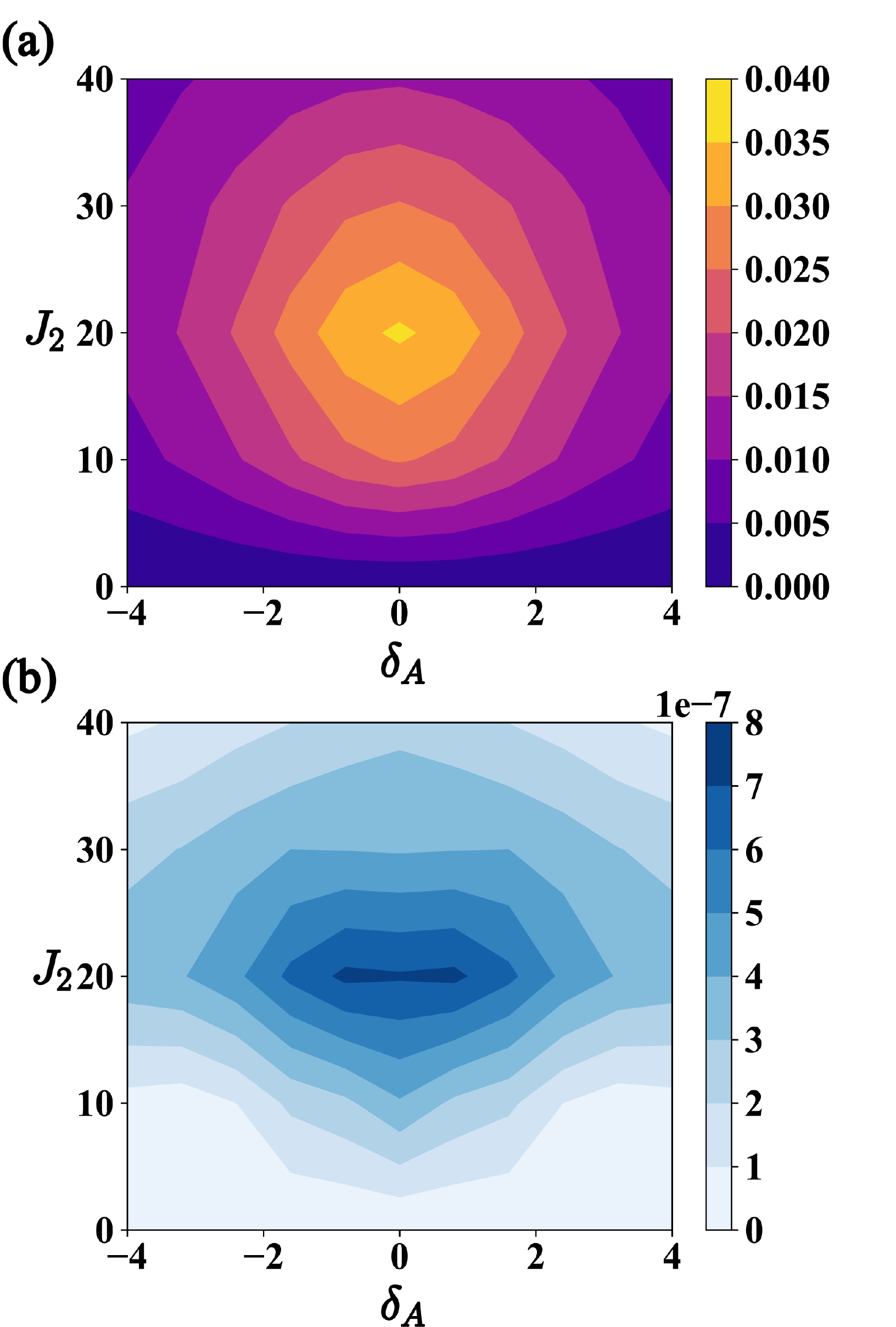}
		\caption{Mutual information $\mathcal{I}\left(\rho_{AE}\right)$ for the steady state of reduced subsystem composed by the agent and the environment subsystems computed by (a) a machine learning protocol including a feedback mechanism and (b) a machine learning protocol without feedback. The system parameters for all the simulations are $\gamma=2\gamma_\phi$, $\delta_R=\delta_E=0$, $J_1=20\gamma$ and $\Omega={10}^{-2}\gamma$. The initial state of the system is given by $|\Psi(0)\rangle=|e\rangle\otimes|g\rangle\otimes|g\rangle$.}
		\label{fig:fig4}
	\end{figure}
	
	Figure (\ref{fig:fig4}a) shows the quantum mutual information $\mathcal{I}\left({\rho}_{{AE}}\right)$ computed with the machine learning protocol depicted in Figure (\ref{fig:fig3}) as a function of the agent qubit detuning $\delta_{A}$ and the coupling between the agent qubit and the qubit register $J_2$. The synchronization region can be described as follows. For weak coupling strength $J_2$, and for any value of the Agent qubit frequency $\delta_{A}$, there are no correlations in the composed system and the mutual information is zero. On the other hand, for small detuning parameter and strong coupling strength $J_2$, the mutual information is different from zero, and both systems (agent and environment) get synchronized. When the agent is highly detuned, the agent and the environment do not correlate for any coupling strength, so the mutual information is zero. Another zone corresponds to the region where the value of $J_2$ exceeds the value of $J_1$. In this case, the increment of $J_2$ implies that both Register and Environment get more correlated than the Agent and Environment subsystems. Thus, by the monogamy relation of Entanglement \cite{68} and Quantum correlations \cite{69}, it is not possible to maximally correlate a third system C if two subsystems A and B are already maximally correlated. Thus, the amount of correlation in the Agent-environment bipartition is less than the amount of correlation in the Register-Environment subsystem, leading to a decreasing of the mutual information $\mathcal{I}\left({\rho}_{{AE}}\right)$.\par
	
	To compare how the feedback mechanism enhances the synchronization between a pair of two-level systems, we compute the mutual information $\mathcal{I}\left(\rho_{AE}\right)$ for the machine learning protocol without feedback (black box in Figure (\ref{fig:fig3})), and compare it with the previous result. In such a case, we observe a synchronization region similar to the obtained with the machine learning protocol with feedback mechanism. However, the mutual information value obtained with this approach exhibits smaller values than the obtained with the protocol including feedback. Thus, our machine learning protocol enhances robustness on the dynamical behavior of the correlations between the Agent-Environment subsystem. 
	Let us study how the quantum reinforcement learning enhances synchronization. Enhance, in this case, means that the numbers of synchronized observables increase after applying the feedback protocol. In such a case, we compute the time evolution for the expectation value of Pauli matrices $\{\sigma_\ell^x,\sigma_\ell^y,\sigma_\ell^z\}$ for both subsystems (A and E) in the situation where agent and environment are equally coupled and detuned with the register qubit. We will compare the result obtained with the machine learning protocol with and without the feedback mechanism.\par
	
\begin{figure}[t!]
		\centering
		\includegraphics[width=1\linewidth]{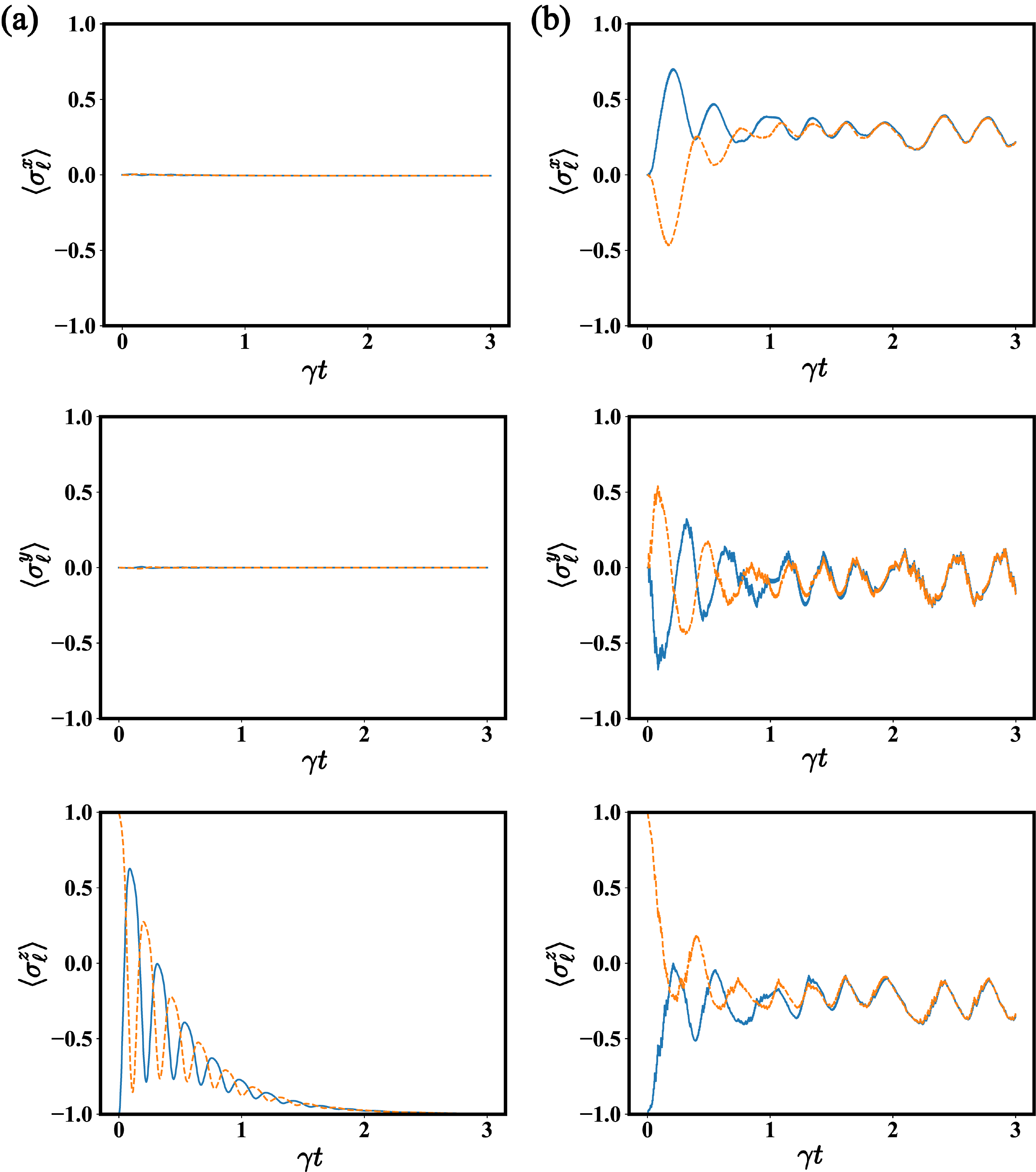}
		\caption{Time evolution of the mean value of the qubits observables. The continuous blue line represents the mean value of Pauli matrices for the agent qubit, and the orange dotted lines show the mean value of the environment qubit observables. Obtained with (a) the digital-analog decomposition without feedback and (b) with the quantum machine learning protocol with reinforcement. The system parameters are $\gamma=2\gamma_\phi$, $\delta_A=\delta_R=\delta_E=10$, $J_1=J_2=20\gamma$ and $\Omega=0.1\gamma$, and the initial state of the system is $|\Psi(0)\rangle=|e\rangle\otimes|g\rangle\otimes|g\rangle$.}
		\label{fig:fig5}
	\end{figure}

	Figure (\ref{fig:fig5}a) shows the qubit observables computed with the ML protocol without the feedback mechanism. As we can see, the observables corresponding to $\sigma^x$ and $\sigma^y$ do not synchronize, while $\sigma^z$ does. This behavior is due to the fact that the gates included in the machine learning protocol produce an effective evolution in the subspace of only one excitation. Then, the observables $\sigma_\ell^x$ and $\sigma_\ell^y$ always vanishes. However, $\sigma_\ell^z$ acting on the state introduces a local phase on the state, then $\left\langle\sigma_\ell^z\right\rangle\neq0$ as shown in the figure. On the other hand, Figure (5b) shows the same observables computed with the ML protocol including feedback. The inclusion of feedback mechanism induces oscillations in the observables $\{\sigma_\ell^x,\sigma_\ell^y\}$, and these observables oscillated with the same frequency, but with a $\pi/2$ phase shift. However, there is a time scale $\kappa t \in\{1,2\}$ in which all observables oscillate with the same frequency and phase. This is due to the change of the register state produced by the feedback mechanism. Moreover, the feedback mechanism induces a change on the system dynamics. Consequently, we observe that the steady-state of the dynamics changes from a completely decayed state to a correlated steady state. Thus, the mutual information embedded in the system after the dynamics is different from zero. In this direction, the robustness exhibited by the correlation agrees with the results obtained in Figure 4 and Figure 5. We observe a small amount of mutual information in the Agent-Environment bipartition (Figure \ref{fig:fig4}a) in the case where the complete set of qubit observables are not synchronized (Fig \ref{fig:fig5}a). Whereas, for the case in which the set of qubits observables synchronizes, (Fig \ref{fig:fig5}(b) machine learning protocol including feedback mechanism) the amount of mutual information $\mathcal{I}\left(\rho_{AE}\right)$ in the Agent-Environment bipartition grows (Fig \ref{fig:fig4}a). Therefore, we conclude that the inclusion of the feedback mechanism leads to an enhancement in the synchronization between two two-level systems. The enhancement relies on the robustness of the mutual information, which acts as a witness for synchronized systems. Furthermore, the proposed protocol increases the number of observables synchronized.
	
	\section{Implementation in superconducting circuits}\label{sec4}
	\begin{figure}[t!]
		\centering
		\includegraphics[width=1\linewidth]{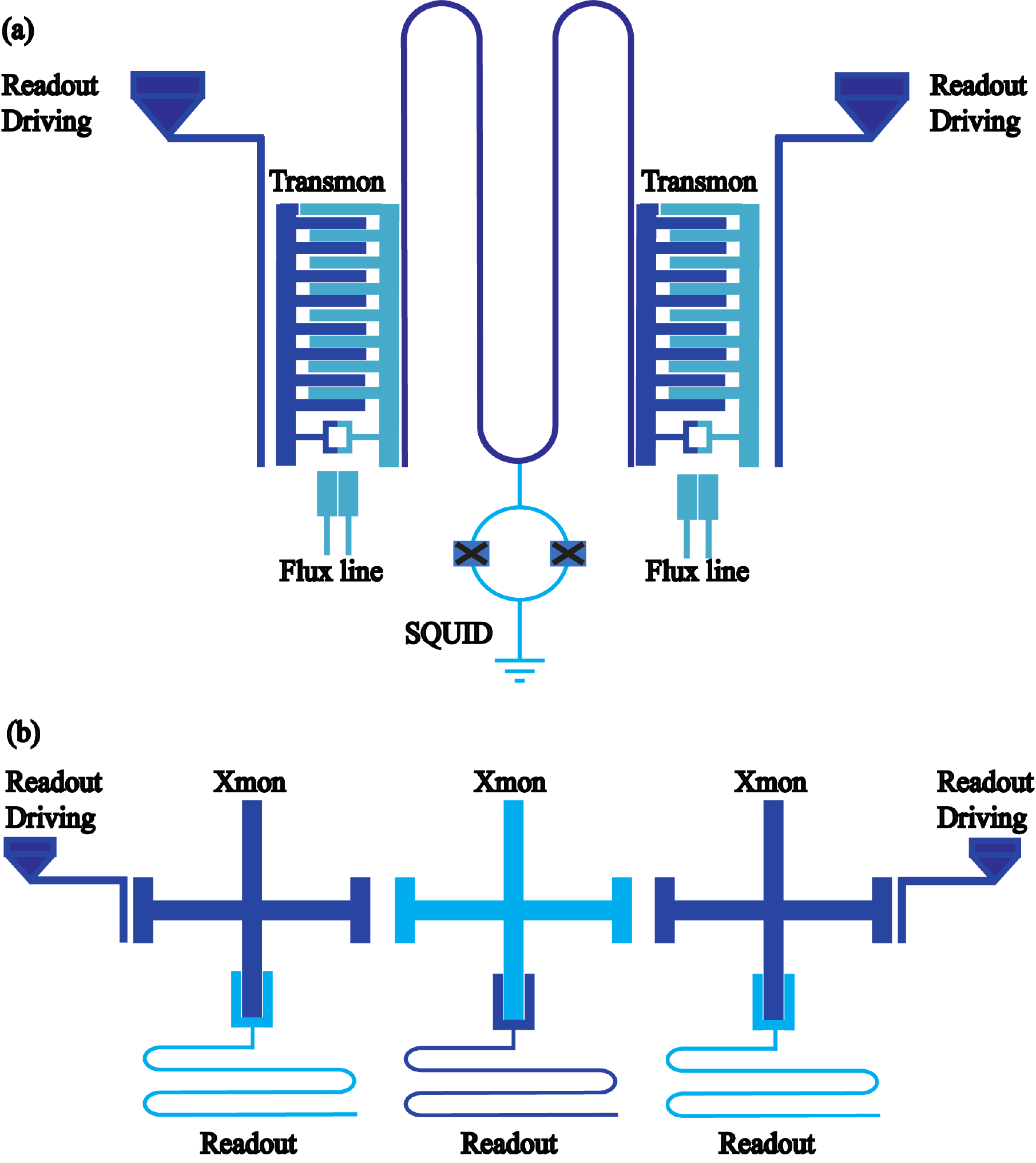}
		\caption{Scheme of the experimental proposal. (a) Two superconducting $\lambda/4$ coplanar waveguide resonators are galvanically coupled to each other through a SQUID. Also, at the edge of each resonator, a transmon device is capacitively coupled to the resonator. (b) Three superconducting Xmon qubits are coupled capacitively to each other.}
		\label{fig:fig6}
	\end{figure}
	Our proposal can be implemented in a circuit quantum electrodynamics architecture with current technology. Indeed, for the realization of the qubit-cavity setup, current technology allows us to connect charge qubits and flux qubits to a microwave transmission line resonator. Our possible setup is composed of two $\lambda/4$ transmission line resonator coupled by a superconducting quantum interference device (SQUID) through the current. This coupling allows us to tune the cavity frequency and the coupling strength between each resonator \cite{70,71,72,73}. Moreover, two transmon qubits \cite{74} are capacitively coupled to the resonator through the voltage at the ends of the transmission line resonator. We choose charge qubits instead of flux qubit because charge qubits show longer coherence times than the flux qubits \cite{61}. For the machine learning protocol implemented with qubits, our proposal based on circuit quantum electrodynamics architecture can be implemented by considering arrays of Xmon qubits \cite{75,76,77}, which offer higher coherence times and fast tunability.
	Current technology has made possible the implementation of quantum feedback in superconducting circuits \cite{78,79,80}. A system based on a closed-loop circuit together with binary measurement has allowed implementing a protocol to reinitialize the state of a qubit, this process is done in a time scale at least one order of magnitude faster than the relaxation time of the two-level system \cite{78,79}. Also, current technology based on transmon qubits coupled to a microwave resonator has allowed to implement weak measurements, which have been empoyed to monitor the Rabi oscillation between qubit states as well as to reconstruct a quantum state \cite{80}.
	
	\section{Numerical convergence digitalized synchronization}\label{sec5}
	\begin{figure}[t!]
		\centering
		\includegraphics[width=1\linewidth]{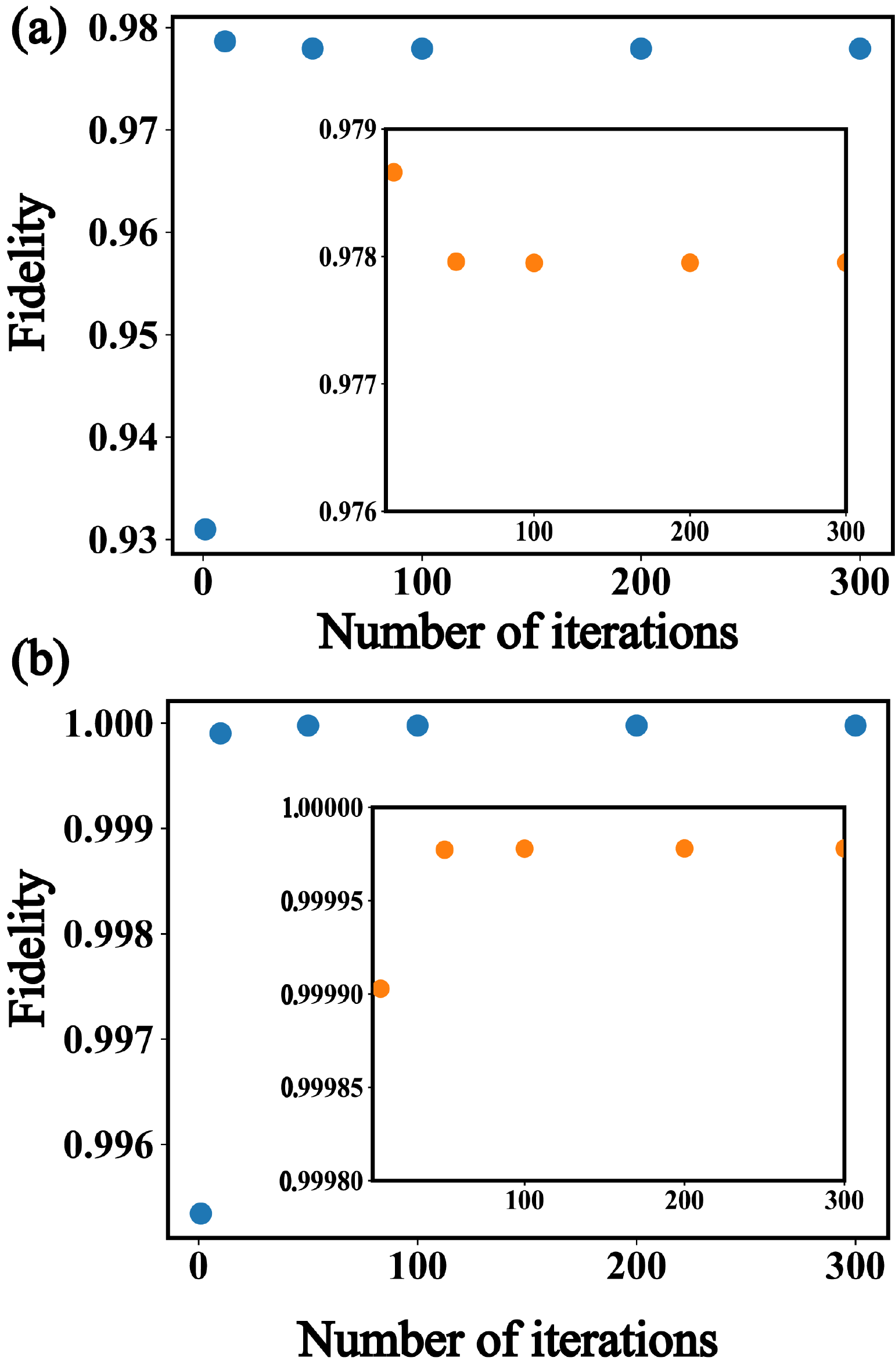}
		\caption{Fidelity $\mathcal{F}$ between the states computed with both approaches for a different number of iteration in a time scale (a) $\kappa t=1$ and (b) $\kappa t=50$. Inside of both plots, there is an enlarged plot showing the plateau obtained for a specific number of iteration}
		\label{fig:fig7}
	\end{figure}
	In this section, we discuss the numerical convergence of the digital-analog decomposition of the master equation proposed in section 2. We also compare the dynamics of the complete master equation with the decomposition, for different time steps $\kappa t$. The comparison between both approaches is performed by computing the fidelity, defined $\mathcal{F}(\rho_{{\rm fully}},\rho_{{\rm digital}}^{n})=({\rm{Tr}}\sqrt{\sqrt{\rho_{{\rm fully}}}\rho_{{\rm digital}}^{n}\sqrt{\rho_{{\rm fully}}}})^{2}$. Here, $\rho_{{\rm fully}}$ is the state obtained by the analog approach, whereas ${\rho}_{\mathbf{digital}}^{n}$ is the state computed with the decomposition for ${n}$ steps. Figure (\ref{fig:fig7}a) shows the fidelity for different number of iteration $n$. For a time ${\kappa t}={1}$, we observe that the fidelity $\mathcal{F}$ exhibits an anomaly. There is an intermediate number of iteration $n=50$ where the fidelity reaches its maximum, and after that we observe a plateau with the optimal fidelity between both approaches. For a time ${\kappa t}={50}$, we observe that, when the number of iterations increases, the fidelity approaches one, and the anomaly is not observed. On the other hand, for $n$ larger than 100, we observe a plateau in the fidelity. Thus, due to all our calculation are  considered for longer ${\kappa t}$,  $n={100}$ subdivisions are sufficient to achieve the optimal fidelity between the states obtained through the full dynamics and the digital-analog decomposition.

	\section{Conclusion}\label{sec6}
	We have shown that quantum synchronization between a pair of two-level systems is achieved by considering the digital-analog decomposition of the master equation which governs the system dynamics. We can identify in this block decomposition the fundamental elements of a quantum machine learning protocol, namely, agent, environment and register. Afterwards, we have also equipped the machine learning protocol with a feedback mechanism based on measurements and reinitialization of the register state together with conditional local operations on the agent and environment subspace, substantially increasing its power and flexibility. Indeed, numerical simulations show an enhancement in the synchronization manifested in two aspects. The first one is the increment of the mutual information between the agent-environment bipartition. The second aspect yields an increasing of operators which synchronize and the time rate in which the synchronization is achieved. Furthermore, by modifying the protocol, we may choose the state in which the system synchronizes. Finally, based on current technologies on superconducting circuit and circuit quantum electrodynamics, we have proposed and implementation of the quantum machine learning protocol with feedback.\par

	\section{Acknowledgments}
	We would like to thank L. Lamata and D. Z. Rossatto for the useful discussions. F.A.C.-L acknowledges support from Financiamiento Basal para Centros Cient\'ificos y Tecnol\'ogicos de Excelencia FB. 0807 and Dirección de Postgrado USACH. J. C. R. acknowledges the support from FONDECYT under grant No. 1140194. M. S. and E. S. acknowledge support from Spanish MINECO/FEDER FIS2015-69983-P and Basque Government IT986-16. The authors also acknowledge the EU Flagship project QMiCS. This material is also based upon work supported by the U.S. Department of Energy, Office of Science, Office of Advance Scientific Computing Research (ASCR), under field work proposal number ERKJ335.

\end{document}